# A Novel Approach to Detect Spam Worms Propagation with Monitoring the Footprinting

Rajesh R Chauhan, M.Tech
*IT Dept, ASTRA,Bandlaguda,*

G S Praveen Kumar, M.Tech
*Associate Professor IT Dept, ASTRA*

**Abstract:-** *One of the key security threats on the Internet are the compromised machines that can be used to launch various security attacks such as spamming and spreading malware, accessing useful information and DDoS. Attackers for spamming activity are volunteer by large number of compromised machines. Our main focus is on detection of the compromised machines in a network that may be or are involved in the spamming activities; these machines are commonly known as spam zombies. Activities such as port scan, DB scan and so on are treated as malicious activity within the network. So to overcome that we develop one of the most effective spam zombie detection system within the network based on the behavior of other systems as if performing the above activities are treated as zombies machines. If any system within the network try's to gather some information about any other system then this is treated as a malicious activity and should be not allowed to do so. SYN packets are used in order to initiate communication within the network so as to establish connection. If any system try's to flood the network with these packets we can make an assumption that the system is trying to gather the information about other system. This is what called footprinting. So we will try to detect any system involved in footprinting and report to the administrator.*

## 1. INTRODUCTION

A computer networks consist of computing devices (network nodes) that pass data to each other along data connections. The connections (network links) between nodes are established using either cable media or wireless media. Network security consists of the provisions and policies adopted by the administrator of the network to prevent and monitor unauthorized access, misuse or modification, or denial of service for a computer in a network. Users of a network can choose or are assigned with an ID and password or other authenticating information that allows them access to information and programs within their authority. Networks can be private, as such used within a company, and others might be open to public access for general usage. In today's era anyone can become a hacker by downloading and using tools available on the Internet. The easiest and effective way of protecting a network from an outside attack is to completely close it off from the outside world. In a closed network only the trusted known parties and sites are provides connectivity; a closed network does not allow a connection to public networks. However, internal threats still exist.

A zombie is a computer connected to the Internet that is compromised, computer virus or trojan horse and can be used to perform malicious tasks of one sort or another under remote direction. Zombie computers forming the Botnet are often used to spread e-mail spam and launch denial-of-service attacks. Footprinting is the technique of gathering information about computer systems and the entities they belong to. Footprinting purpose is to learn as much as you can about a system, its access capabilities, its services and ports, and the security aspects. To perform a successful hack on a system, it is preferred to know as much as you can, even a little if not everything, about that system. If you can gather some information about a system, the company that owns that system, with the right personnel, can find out anything they want about you.

## 2. LITERATURE REVIEW

A literature review is a text written by someone to consider the critical points of current knowledge including substantive findings, as well as methodological and theoretical contributions to a particular topic. A literature review is most important to collect the information of the methods and system prior developed which can help in giving a brief idea of what parameters we need to keep in mind and what we need to develop in order to make our project more efficient than the rest.

2.1 BotSniffer: Detecting Botnet Command and Control Channels in Network Traffic: One of the most serious security threats are botnets, possessing the command and control (C & C) channel characteristics that uses protocol such as IRC, HTTP which creates the problem in detection of botnet (C&C). Because of the pre-programmed activities related to C&C, bots in same botnet have spatial-temporal correlation and similarity. BotSniffer capture the spatial-temporal correlation in network traffic and utilize statistical algorithms to detect botnets with boundaries of false positive and false negative rates and thus can be use in real-world network traces. A botnet C&C channel has a "botmaster" to direct the actions of bots in a botnet. If C&C channel is interrupted or broken the botmaster will be unable to control botnet. So we need to understanding and detecting the C&Cs communication channel in the botnet. In order to control centralized botnet C&C we need to study two styles. The first is the "push" style, where commands are pushed or sent to bots i.e. IRC (Internet Relay Chat) Based. The second is the "pull" style, where commands are pulled or downloaded by bots i.e. HTTP-based. Secondly, anomaly-based detection algorithms to identify both IRC and HTTP based C&Cs in a port independent manner. Advantage of this is that no prior knowledge of C&C servers or content signatures is needed, encrypted C&C can be detected, no need of large number of bots and can detect botnet with a single member in the monitored network. Thirdly, *BotSniffer*, based on anomaly detection algorithms which can be used in real world network trace.

2.2 DMTP: Controlling Spam through Message Delivery Differentiation: Spamming activities i.e unsolicited commercial email has become a big problem in today's internet world. Generally sender-driven email system lacks receiver control over email delivery. In Differentiated Mail Transfer Protocol (DMTP) receivers are given greater control over delivery of messages from different classes of senders on the Internet. DMTP provides several important advantages in controlling spam: retrieval behavior of receivers is determined by the delivery rate of spam rather than being controlled by spammers. Spammers are made to stay online for much longer periods of time. No extra effort is needed for regular correspondents to communicate with the receiver. Asynchronous messages like email are delivered on the Internet primarily using two different models: sender-push and receiver-pull.

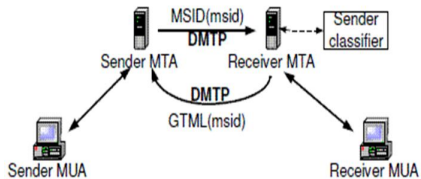

Fig 2.2: DTMP based email system

In the sender-push model, delivery of traffic is controlled by the sender, and whatever the senders push is passively accepted by receiver .The receiver-pull model the receivers can control what they want to retrieve data from the senders. DMTP uses the algorithm where receiver classifies each sender in three classes i.e well known spammers, regular contacts and unclassified sender and accordingly accept or rejects the mails from the sender.

2.3 Understanding Localized-Scanning Worms: Vulnerable hosts in the local network can be searched by attackers by using a simple technique called localized scanning. Localized scanning trades off between the local and the global search of vulnerable hosts and has been used by Code Red II and Nimda worms. As such a strategy is so simple yet effective in attacking the Internet, it is important that defenders understand the spreading ability and behaviors of localized-scanning worms. We first characterize the relationships between vulnerable-host distributions and the spread of localized-scanning worms through mathematical modeling and analysis, and compare localized scanning with random scanning. Self-propagating Internet worms have posed significant threats to network security. For example, Code Reg, Nimda, and Witty worms infected hundreds of thousands of computers and cost tremendous efforts to eliminate them. Therefore, it is important that we understand how worms spread to design effective countermeasures accordingly. A worm spreads by using distinct scanning mechanisms including topological and hit list scanning. Our focus, however, is only on scanning worms that probe the entire IPv4 address space or the routable address space, such as random, routable, importance, and localized scanning. *Random scanning* chooses target IP addresses at random and is exploited by Code Red and Witty worms. *Routable scanning* selects targets only in the routable address space by using the information provided by BGP routing table. *Importance scanning* exploits an uneven distribution of vulnerable hosts and focuses worm scans on the most relevant parts of the IPv4 address space. In this work, our focus is on localized scanning, which has been used by such famous worms as Code Red II and Nimda.

## 3. IDENTIFING THE FLOODING & SPAMS

Firewalls follow some simple rules such as to deny or allow protocols, IP address or port. Some attacks such as DoS attacks are too complex even though for today's firewalls, e.g. suppose there is an attack happening on port 80 (web service), the attack cannot be prevented by firewall because they doesn't have the capability to differentiate between good traffic from DoS attack traffic. Routers which are places in between the network that may be affected even before the firewall gets the traffic.

Flooding is a kind of Denial of Service (DoS) attack that is designed in order to bring a network or service down by flooding it with huge amounts of traffic. Flood attacks take place when a network or service becomes too weighted with packets initiating incomplete connection requests that restricts genuine connection requests. Establishment of a new connection is indicated by SYN packet. In TCP connection the first packet that is sent across is known as a "SYN" or "synchronize" packet. For example, when we try to connect to http://www.google.com, the first packet send out of your systems will be a SYN packet to the HTTP port 80 on www.google.com. Your browser is notifies the web server that it wants to connect. A SYN flood is a form of denial-of-service attack in which an attacker sends a succession of SYN requests to a target's system in an attempt to consume enough server resources to make the system unresponsive to legitimate traffic.

Normally for a TCP connection a series of messages normally flow between client and server like this: The client requests a connection by sending a SYN (*synchronize*) message to the server. The server in response *acknowledges* this request by sending SYN-ACK back to the client. The client in response to SYN-ACK sends ACK, and then only the connection is established. This sending of packets from client to server and vice versa is called the TCP three-way handshake, and this is how the foundation is set for every connection established using the TCP protocol. A SYN flood attack can be done by not responding to the server with the respective ACK code. The malicious client or the compromised client can either simply not send the expected ACK, or by spoofing the source IP address in the SYN i.e acting as a legitimate user, make the server to send the SYN-ACK to a falsified IP address - which in turn will not send an ACK because it "knows" that it never sent a SYN. This *half-open connections* will occupy all the resources on the server so that no new connections is possible which results in a denial of service (DoS) for the legitimate traffic. In addition to the above description worms are also one of the greatest threats today. A network worm is a standalone program that tries to copy itself to other computers or nodes connected to the same LAN (Local Area Network). A network worms travel from one computer to another using some kind of share media. The group of such infected machines is known as botnet. A botnet is a collection of robots, or software agents, that run autonomously and automatically. A botnet operator sends out viruses or worms, infecting ordinary users computers, and carry payload as a malicious application—the bot. The bot logs into a particular C&C server from the infected PC. A service of the botnet is purchased from the operator by a spammer. The spam message is provided by the spammer to the operator, who via the IRC server instructs the compromised machines and causing them to send out spam messages. Spam is flooding the Internet with same message having multiple copies, forcing the message on people who would not otherwise choose to receive it. Spams are mainly of two types, and have different types of effects on Internet users. Cancellable Usenet spam is a single message sent to 20 or more Usenet newsgroups. Email spam targets individual users with direct mail messages. Email spam lists are mostly created by scanning Usenet postings carried out by stealing Internet mailing lists, or searching for addresses on the Web. One particularly nasty variant of email spam is sending spam to mailing lists. Because many mailing lists limit activity to their subscribers, spammers will use automated tools to grab as many mailing lists as possible, so that they can get the lists of addresses, or use the mailing list as a direct target for their attacks.

## 4. FOOTPRINTING AND SYN FLOOD ATTACK DETECTION

Detection System Sub:
1. Packet Capturing.
2. Field Extraction.
3. Field Information Storing.
4. SYN-Flood attack detection.
5. Footprinting attack detection.

4.1 Packet Capturing: This module captures or records all the packets which are flowing through the IDS device. Packet capture related to computer networking is a term for retrieving a data packet moving over a computer network. After a packet is captured, it is needed to store temporarily so that it can be analyzed.

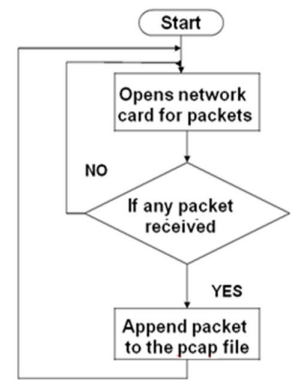

Fig 4.1: flow chart of packet capturing

Packets running over a network are capture and examine in real-time running by using different packet capturing techniques. Among them filtering is one kind, in which filters are made to pass over network nodes or devices where data from them is captured. Rather than filtering just a specific portion of a packet, complete packets can also be captured. Complete packet includes two things: a payload along with a header. The actual content of the packet is in payload while the extra information is in header part, including the packet's source and destination address.

**4.2 Field Extraction:** Field Extraction refers both to the process by which fields are extracted from event data, and the results of that process.

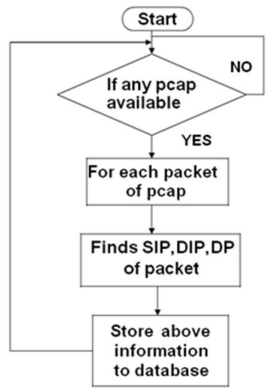

Fig 4.2: flow chart of field extraction

Field extraction either can take place before events are indexed or after event indexing. In this module we extract all the fields present in the network. This module reads the pcaps that are stored by the Packet capturing module. Opens each pcap, and for the each packet extracts, source IP(SIP), Destination IP(DIP), Source port(SP), Destination Port(DP), and Packet TCP-Flag (Ex: SYN, FIN, RST) and passes these values to the next module. This process will be repeated for all the packets of all the pcaps.

**4.3 Field Information Storing:** This module inserts newly arrived packet information i.e extracted from the above module into a new file. If PCF module processed the entries deletes them from the database.

**4.4 SYN-Flood Attack Detection:** Partial Completion Filter (PCF) consists of parallel stages each containing hash buckets that are incremented for a SYN and decremented for a FIN.

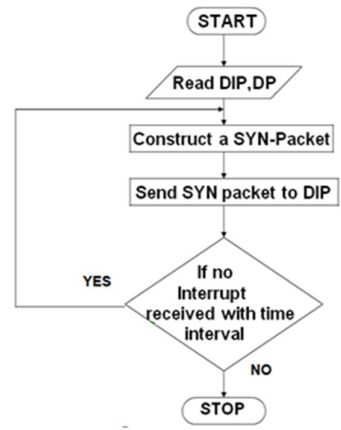

Fig 4.3: flow chart of SYN flood detection

In this module we use PCF for identifying the SYN packets in the network. PCF for SYN-Flood attack can be designed as PCF (SYN/FIN, any SIP, any SP, same DIP, same DP, 512). First part of the PCF indicates, If the Packet is of type SYN/FIN packet information will used for the processing, If the packet is of type SYN logical counter will be incremented. If packet is of type FIN logical counter will be decremented, Alerts will be raised only when the logical PCF counter reached to 512.

**4.5 Footprinting Attack Detection:** A foot print or port-scan is an attack that sends requests from a client to a range of server port addresses on a host, with the intention of finding an active port and exploiting a known vulnerability of that service.

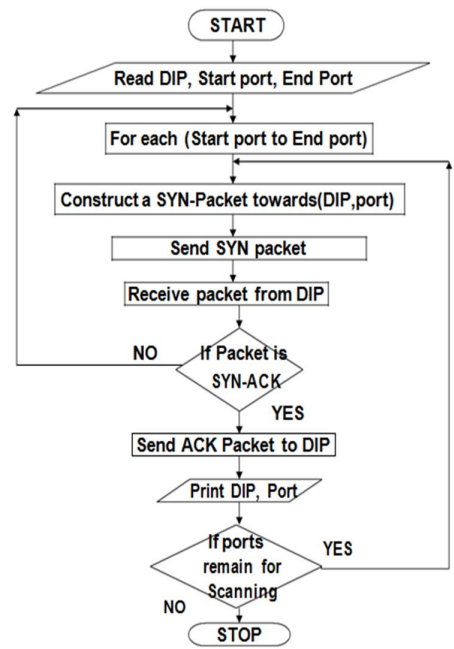

Fig 4.4: flow chart of footprinting attack detection

In this module we use PCF for identifying the port-scan in the network. This attack program should send TCP packets to the target system and should display the open ports on the target system. We first start sending the SYN packets towards first destination based on the destination IP and port and then continue it to rest of the systems in the network. Then the packet in response forwarded by the destination are extracted for the information such as destination port, source port and destination IP,SYN packet, FIN, packet, ACK packet and so on and then this information is stored in files. Then we calculate the SYN packets toward specific destination IP on to different ports as well as the ACK received in response and if this value reaches to threshold value then we assume that there has been footprinting done on that IP and so we construct the report about that and present it to the administrator so that he can take specific action on that.

### 5. SYSTEM ARCHITECTUTE

System architecture is the conceptual model that defines the behavior, the structural and over all views of a system. An architecture description is a formal description and representation of a system, organized in a way that supports reasoning about the structures of the system.

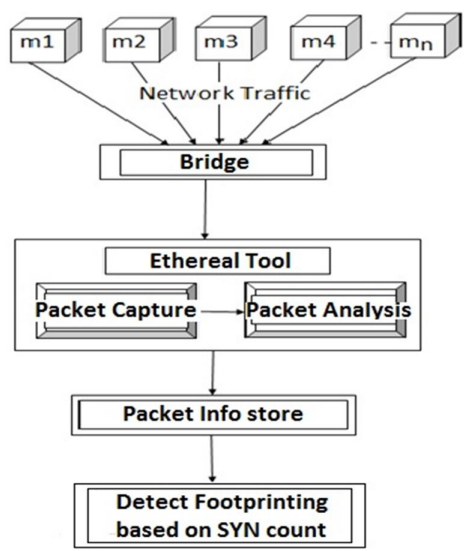

Fig 5.1: system architecture

The system architecture above shows the systems m1, m2, -- mn collectively making a network. The network traffic passes through the bridge in which we can place the detection system or it can be places in any of the systems within the network. Bridges inspect incoming traffic and decide whether to forward or discard it. The network traffic after coming to a bridge is the passed to the tethereal toll. It captures packet data flowing in a live network, or can read packets previously saved in a capture file, either can writing the packets to a file or print a decoded form of those packets to the standard output. The packets captured are then analyzed in order to get information from the packets. Packet analysis means capturing the packets and then converting it to raw data into readable format. The information extracted is then saved in a file having all the information of the packet, from that 5 tupple information is extracted which is needed for the detection system. Then the input is given to the detection system where the information extracted is placed against the footprint detection logic of syn count again and again in regular interval of time and based on that the system to be attacked is detected and then a log is prepared and then this log is presented to the user of the system so that specific action can be carried out.

### 6. PACKETCAPTURE, FIELD EXTRACTION

In order to avoid any kind of attack happening on the system or within the network we firstly need to know what kind of attack is being happening and from where that attack is been carried and finally on which system this attack is being done. In order to find answer of all this questions we need to collect the information being flowing in the form of packets between the systems of the network. So we need to capture those packets and then extract that information from the packets. We use a VMware workstation in-order to create a bridge network in to a single pc so that we can simulate the project properly. BTLinux act as an attacker machine which sends some SYN packets to the detection system in the bridge network. The detection system capture's the packet's being send by the BTLinux and then analyze them to detect attack. Firstly it is highly impossible to see the information flowing within the network by naked eye and secondly if it possible we cannot understand anything from that. So to do this work properly and efficiently we are using a tool i.e tethereal for this purpose. Tethereal - Dump and analyze network traffic. It lets you capture packet data from a live network, or read packets from a previously saved capture file, either printing a decoded form of those packets to the standard output or writing the packets to a file.
Packets Captured: We initiate tethereal by the program by using system command. It takes some parameters which are used to set the network interface, the client, single line packet writing in a specific file, set the duration of packet capture and so on.
Field Information Extraction**:** After capturing the packets the information is to be extracted i.e field information extraction is done. First we check whether any lines are present in the file or not and if any are present then they are copied to a file. Field Information Storing: Here parsing of packets is done i.e dividing of packets into parts and respective information is stored in source IP SRCIP, destination IP DEST, and source Port SRC PORT and so on. Detecting Footprinting Attack: After extracting the information from the packets the logic is set so as to detect any kind of malicious activity is going on, in our case we need to detect whether footprinting is being done or not. In order to detect footprinting we need to keep track if the SYN packets being send to the different port for same destination IP i.e SYN flood is being done. If the threshold value of the SYN flood is excided the we can make an assumption that there is an footprinting activity going on and thus we can report the administrator about his and show the port and destination IP at which footprinting is being done as shown below.

## 7. CONCLUSION

As we have seen earlier in there have been many suggestions given in order to detection of spam in the network and how to avoid or restrict the attacks happening on the network systems. Previously attack detection was done after the attack was happened and that mostly outside the local network, based on characteristics of their behaviors and so on. There were some or the other restrictions and limitations of the other systems mentioned previously. In our system we have tried to overcome some of the limitations of other systems. We are able to detect the attack before it is actually happened based on the initial packet being received within the network and this can be a big advantage. The boundaries such as false positive and false negative have been avoided so as not to restrict the system between them. The threshold value of the syn count can be set as par the network so that there can be affective detection. So examining the above points we conclude that this system is having more advantages as compared to all other system we have studied and is effective in prevention and detection of attack being done within a network. The snap short of result is as shown below

**Table of Alerts**

| Log Details |
|---|
| Spam/Worm Affected IP: 192.168.1.100 AttckName:FOOT PRINTING ATTACK Detected No.of.Scans:4 |
| Spam/Worm Affected IP: 192.168.1.100 AttckName:FOOT PRINTING ATTACK Detected No.of.Scans:4 |

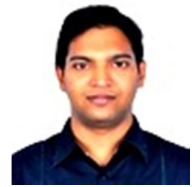

First Author: Rajesh R Chauhan received his Bachelor of Engineering (BE) degree from Sant Gadge Baba Amravati University in 2011. He is currently pursuing M. Tech. in Information Technology, from ASTRA, Bandlaguda affiliated from Jawaharlal Nehru Technological University, Andhra Prades. His research interests are in Information and Network Security, Cloud Computing, Cryptography, Digital Signal Processing and Image Processing.

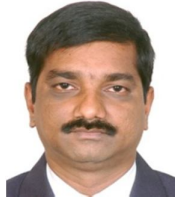

Second Author: G S Praveen Kumar, Associate Professor, Aurora Scientific Technological and Research Academy, completed his M. Tech ( CSE) and pursuing Ph.D in Network Security from SV University, Tirupati. His areas of research are Network Security and Fuzzy logic and published two papers in International journals. He participated in an international conference in network security in Singapore. Currently he is working in Masquerader Detection System for Graphical User Interface using fuzzy logic.